\documentclass[twocolumn,showpacs,preprintnumbers,aps,pra]{revtex4-1}

\usepackage{hyperref}
\hypersetup{
    backref =       true,
    pagebackref  =  true,
    colorlinks =    true,
    linkcolor =     [rgb]{0.0,0.0,0.8},
    anchorcolor =   [rgb]{0.0,0.0,0.8},
    citecolor =     [rgb]{0.0,0.0,0.8},
    filecolor =     [rgb]{0.0,0.0,0.8},
    urlcolor =      [rgb]{0.0,0.0,0.8},
    pdftitle=       {Title},
    pdfsubject=     {Title},
    pdfauthor=      {A. Uthor}
}

\usepackage[utf8]{inputenc}
\usepackage{amsmath}
\usepackage{graphicx}

\makeatletter

\@ifundefined{textcolor}{}
{
 \definecolor{BLACK}{gray}{0}
 \definecolor{WHITE}{gray}{1}
 \definecolor{RED}{rgb}{1,0,0}
 \definecolor{GREEN}{rgb}{0,1,0}
 \definecolor{BLUE}{rgb}{0,0,1}
 \definecolor{CYAN}{cmyk}{1,0,0,0}
 \definecolor{MAGENTA}{cmyk}{0,1,0,0}
 \definecolor{YELLOW}{cmyk}{0,0,1,0}
}

\usepackage{txfonts}
\usepackage{dcolumn}
\usepackage{bm}
\usepackage{natbib}
\usepackage{siunitx}
\usepackage{microtype}
\makeatother

\begin{document}

\title{Optomechanically induced carrier-envelope-phase dependent effects and their analytical solutions}

\author{Jinyong Ma$^{1,2}$, Jinghui Gan$^{2}$, Giovanni Guccione$^{1}$, Geoff T.\ Campbell$^{1}$, Ben C.\ Buchler$^{1}$, Xinyou L\"{u}$^{2}$, Ying Wu$^{2}$, and Ping Koy Lam$^{1}$}\email{Ping.Lam@anu.edu.au}

\affiliation{$^1$Centre for Quantum Computation and Communication Technology, Department of Quantum Science, Research School of Physics and Engineering, The Australian National University, Canberra ACT 2601, Australia}
\affiliation{$^2$School of Physics, Huazhong University of Science and Technology, Wuhan 430074, China}

\date{\today}
\begin{abstract}
To date, investigations of carrier-envelope-phase (CEP) dependent effects have been limited to optical pulses with few cycles and high intensity, and have not been reported for other types of pulses. Optomechanical systems are shown to have the potential to go beyond these limits. We present an approach using optomechanics to extend the concept of the traditional CEP in the few-cycle regime to mechanical pulses and develop a two-step model to give a physical insight. By adding an auxiliary continuous optical field, we show that a CEP-dependent effect appears even in the multi-cycle regime of mechanical pulses. We obtain the approximated analytical solutions providing full understanding for these optomechanically induced CEP-dependent effects. In addition, our findings show that one can draw on the optomechanical interaction to revive the CEP-dependent effects on optical pulses with an arbitrary number of cycles and without specific intensity requirements. The effects of CEP, broadly extended to encompass few- and multi-cycle optical and mechanical pulses, may stimulate a variety of applications in the preparation of a CEP-stabilized pulse, the generation of ultrasonic pulses with a desired shape, the linear manipulation of optical combs, and more. 
\end{abstract}

\maketitle
\section{Introduction}
The simple premises behind an optomechanical system (OMS), where a mechanical resonator interacts with an optical field via radiation pressure force, offer untold opportunities that were even hard to imagine until recent years. The rapid development of optical control in optomechanical devices \cite{weis_OMIT_opto, safavi-naeini_slow-light_opto, nie_low-excitation_opto, jiang_absorption-mech-drivingt_opto} brings applications of OMS to the cutting edge of metrology, such as precision measurements of acceleration~\cite{krause_accelerometer_opto}, magnetic fields~\cite{forstner_magnetometer_opto}, weak forces~\cite{gavartin_force-measure_opto}, and electrical charges~\cite{zhang_charge-measure_opto}. New frontiers are also targeted for position measurements, with pioneering research aiming to push the precision towards the standard quantum limit~\cite{clerk_quantum-noise-measurement_review}. In fact, the applications of OMSs in the field of metrology are too many to be listed comprehensively, and there is always potential for further extension to other significant physical quantities. The carrier-envelope phase (CEP) of a given pulse is one example. 

The CEP is the phase between the envelope of a pulse and the carrier frequency at which the pulse is modulated. In the field of ultrafast optics, the rapid progress of CEP measurement techniques has made it possible to fully manipulate laser pulses, paving the way to numerous novel physical phenomena and applications. Some of these include, for example, super high-resolution measurements on atomic and molecular systems~\cite{drescher_time-resolved_pulseapp}, efficient tomography of molecular orbitals~\cite{itatani_tomographic_pulseapp}, and generation of intensive femtosecond electron beams \cite{mangles_electron-beam_pulseapp, faure_plasma-electron-beams_pulseapp,geddes_electron-beam_pulseapp}. Despite their manifest popularity, so far CEP-dependent effects have mostly been limited to pulses with few oscillations within the envelope (few-cycle pulses), and the measurement techniques involved result in impossible for pulses whose intensity is too low to excite tunneling ionization~\cite{nakajima_CEP-weak_opulse, wu_weak-field_CEP}. Sparked by the growing importance of CEP measurements, recent works extended the exploration of CEP effects towards multi-cycle~\cite{sansone_multi-cycle-CEP_opulse} and low-intensity~\cite{nakajima_CEP-weak_opulse} regimes, finding interesting applications in the generation of CEP-stabilized ultra-short laser pulses~\cite{tzallas_many-CEP_opulse, tang_multi-cycle_opulse} and the coherent control of molecular and electron collisions~\cite{tzallas_absolut-CEP-multi-cycle_opulse}. The CEP-dependent effects of laser pulses with both an arbitrary number of cycles and arbitrary intensity are, nevertheless, still an open issue that needs to be explored.

It is also worth noting that almost all the current research on CEPs are focused only on pulses of an optical nature, missing the potential significance of CEPs for other types of pulses. The generation and manipulation of ultrasonic pulses are currently topics of intense research~\cite{temnov_ultrafast_sound, obrien_ultrafast-nano_sound, saito_pico-nickel_sound, pezeril_generation_sound}, accounting for a number of recent achievements such as super-resolution imaging~\cite{clark_fast-imaging_soundapp, errico_ultrafast-imaging_soundapp}, control of microfluidics~\cite{friend_microfluidics-sound_review}, and coherent magnetization precession~\cite{scherbakov_magnetization_soundapp}. Remarkable advancements in the generation, detection, and control of magnetic~\cite{gerrits_ultrafast-generation_megnetic-pulse} and electric~\cite{lampin_detection_electricalpulse, keil_generation_electricalpulse} pulses are also pushing for the development of new applications~\cite{gerrits_ultrafast-generation_megnetic-pulse, back_magnetization-reversal_magnetic, nuccitelli_self-destruct_electricapp}. In view of the compelling demand, an extension of CEP measurement and control techniques to new types of pulses is very appealing. From this perspective, the diversity of experimental setups in optomechanics~\cite{aspelmeyer_cavity-opto_review} makes an OMS become an incredibly useful platform that can act as transducers between different types of pulses generated in various systems. For example, ultrasonic pulses can be transformed into mechanical pulses by the acoustic radiation pressure force generated by the ultrasound~\cite{friend_microfluidics-sound_review}, while electric or magnetic pulses can be changed to mechanical actuation by piezoelectric \cite{xu_controllable-mechanical-drive_opto} or piezomagnetic effects. This indicates that, in principle, one can focus the study of CEP measurements and control of a pulse on a generic mechanical pulse while at the same time benefiting from the profusion of applications associated with pulses of different natures.

In this work, we present a model addressing the CEP-dependent effects for both mechanical pulses and optical pulses using optomechanics. Even though several theoretical pioneering works \cite{roudnev_CEP-theory_opulse, schafer_three-step_hhg, corkum_plasma_three-step_hhg, milosevic_theory-few_CEP} in the field of ultrafast optics have qualitatively and quantitatively described the CEP-dependent effects appearing in atoms or molecules, the analytical description and physical insight of CEP-dependent effects emerging in an optomechanical system are still not well understood. We set down the approximated analytical results that are strongly supported by accurate numerical simulations, leading us towards a clear physical picture of optomechanically induced CEP-dependent effects. In the resolved-sideband regime where the resonance frequency of the mechanical oscillator is much larger than the linewidth of the cavity, our analysis starts from few-cycle mechanical pulses, showing that the plateau width (PW) of the cavity output spectrum directly depends on the CEP. In the multi-cycle regime of mechanical pulses, where the traditional CEP-dependent effects are washed out, we introduce an auxiliary continuous field to reveal a novel CEP-dependent effect: the plateau height difference (PHD) between the Stokes and the anti-Stokes sidebands in the spectrum. The outcomes are easily adapted to the case of optical pulses with an arbitrary number of cycles and without specific intensity requirements. As a consequence, CEP measurements are extended to the domain of ultra-weak, multi-cycle optical pulses, as well as all other types of pulses that an OMS can interact with. Along lines similar to the applications of the CEP of optical pulses, the CEP measurement of mechanical pulses has the potential to be applied to the shape detection and manipulation of ultrasonic, electric, or magnetic pulses.

\section{CEP-dependent effect in the few-cycle regime}
\begin{figure}[htb]
\includegraphics[width=0.48\textwidth]{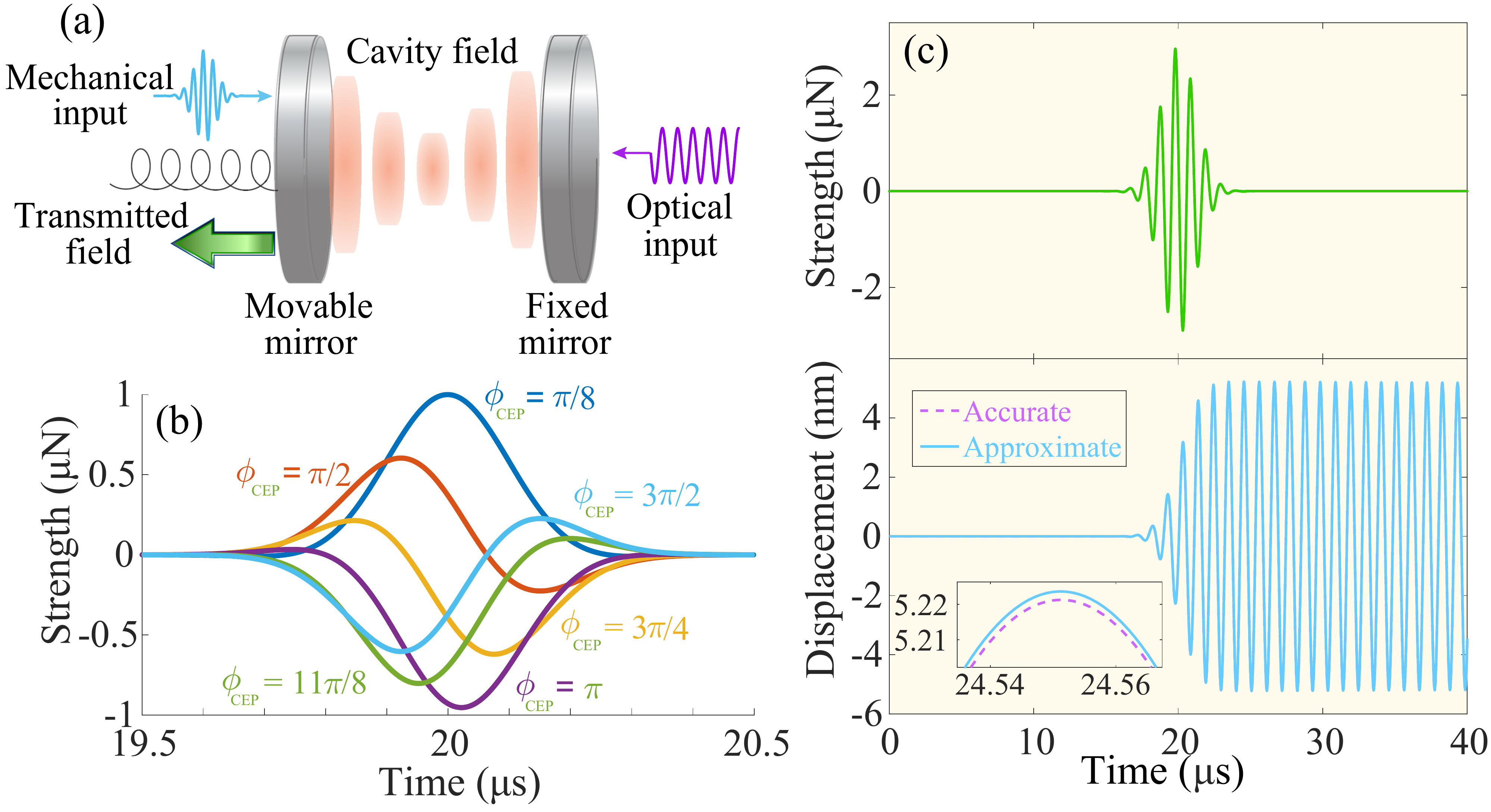} \caption{(a) Schematic diagram of the optomechanical system. \;(b) Different shapes of pulsed mechanical forces depending on the CEP. We set $A=\SI{1}{\micro\newton}$ and $t_\textrm{p}=\SI{0.2}{\micro\second}$. \;(c) The dynamical evolution of a mechanical pulse and the corresponding mechanical displacement. The other relevant parameters used in this work are borrowed from a recent state-of-the-art experiment~\cite{groblacher_strong-coupling-exp_opto}: $\gamma_\textrm{m}=2\pi\times\SI{140}{\hertz}$, $\kappa=2\pi\times\SI{215}{\kilo\hertz}$, $\omega_\textrm{m}=2\pi\times\SI{947}{\kilo\hertz}$, $m=\SI{145}{\nano\gram}$, $g_0=2\pi\times\SI{2.7}{\hertz}$, $G=-g_0/\sqrt{\hbar/(m\omega_\textrm{m})}$.}
\label{fig:1}
\end{figure}

We consider a cavity OMS [see Fig.~\ref{fig:1}(a)] consisting of a fixed mirror and a movable mirror (i.e., mechanical resonator), with characteristics similar to a recent pioneering experimental scheme~\cite{groblacher_strong-coupling-exp_opto}. The system is driven by a continuous optical input with frequency $\omega_\textrm{l}$ and an external mechanical pulse with carrier frequency $\omega_\textrm{m}$, and its Hamiltonian in the rotating frame of $\omega_\textrm{l}$ is
\begin{align}
H=\left(\frac{\hat{p}^{2}}{2m}+\frac{m\omega_\textrm{m}^{2}\hat{q}^{2}}{2}\right)-\hbar\Delta\hat{c}^{\dagger}\hat{c}+\hbar G\hat{q}\hat{c}^{\dagger}\hat{c}+H_\textrm{md}+H_\textrm{od},	\label{eq:Ha}
\end{align}
with $H_\textrm{md}$ and $H_\textrm{od}$ being respectively the mechanical drive and optical drive, which are chosen as in this section: $H_\textrm{md}=A\exp\!\left[-2\ln2\left(\frac{t-t_{0}}{t_\textrm{p}}\right)^{2}\right]\cos\left(\omega_\textrm{m}t+\phi_\textrm{CEP}\right)\hat{q}$ and $
H_\textrm{od}=i\hbar\left(\varepsilon_\textrm{l}\hat{c}^{\dagger}-H.c.\right)$. Here, $\hat{q}$ and $\hat{p}$ are respectively the position and momentum operators of the mechanical resonator of mass $m$, whereas $\hat{c}$ ($\hat{c}^{\dagger}$) is the bosonic annihilation (creation) operator of the optical mode. For simplicity, we consider the case that the mechanical pulse matches the mode of the mechanical resonator by assuming the mechanical eigenfrequency to be $\omega_\textrm{m}$. 
The parameter $\Delta=\omega_\textrm{l}-\omega_\textrm{c}$ is the cavity detuning (where $\omega_\textrm{c}$ denotes the cavity resonant frequency), and $\kappa$ is the cavity decay rate. The input field inside the cavity $\varepsilon_\textrm{l}$ is obtained using $\varepsilon_\textrm{l}=\sqrt{2\kappa P_\textrm{l}/(\hbar\omega_\textrm{l})}$ with $P_\textrm{l}$ being the input laser power. The term $\hbar G\hat{q}\hat{c}^{\dagger}\hat{c}$ describes the optomechanical interaction, with the coupling strength $G$. For the input pulse we consider, in a representative form, a Gaussian envelope such as the ultrasonic pulse discussed in Ref.~\cite{sakadzic_shape-gaussian_sound}, with $A$ being the amplitude, $t_\textrm{p}$ being the full width at half maximum and $t_0$ being the arrival time of the center point, set to $\SI{20}{\micro\second}$ in all the following simulations. The parameter $\phi_\textrm{CEP}$ represents the carrier-envelope phase, whose importance in determining the pulse shape [see Fig.~\ref{fig:1}(b)] is essential for the control of a few-cycle pulse. In obtaining the dynamics from the Hamiltonian we ignore the quantum fluctuations~\cite{carmon_chaotic_opto, weis_OMIT_opto, bai_OMIT-quadratical_opto}, and consider the classical Langevin equations as follows by adding the mechanical ($\gamma_\textrm{m}$) and optical ($\kappa$) dissipation terms:

\begin{align}
\ddot{q}+\gamma_\textrm{m}\dot{q}+\omega_\textrm{m}^{2}q&=-\frac{\hbar G}{m}\left|c\right| ^{2}+\frac{A}{m}e^{-\beta^2\left(t-t_{0}\right)^{2}}\cos(\omega_\textrm{m}t+\phi_\textrm{CEP}), \label{eq:dif1}\\
\dot{c}&=-\left[\kappa+i\left(-\Delta+Gq\right)\right]c+\varepsilon_\textrm{l}. \label{eq:dif2}
\end{align}
where $\beta=\sqrt{2\ln2}/t_\textrm{p}$. Since both differential equations are non-linear, an analytical solution to both is non-trivial. However, we shall derive the approximated analytical solutions if we consider a \textit{strong-drive approximation}, where the drive to the mechanical motion is typically much more intense than the effect of the radiation pressure force, and thus we are able to drop the term $-\frac{\hbar G}{m}\left|c\right| ^{2}$ from Eq.~(\ref{eq:dif1}). Ignoring the effects of light onto the mechanics does not completely eliminate the optomechanical interaction: the position of the mirror still exerts its influence on the cavity field, acting as a `transducer' for the mechanical pulse. For the subsequent analysis we focus on two time windows, one during which the pulse is interacting with the system and the other starting right after the end of the mechanical pulse and ending before the damping of the displacement becomes appreciable.

\begin{figure}[htb]
\includegraphics[width=0.49\textwidth]{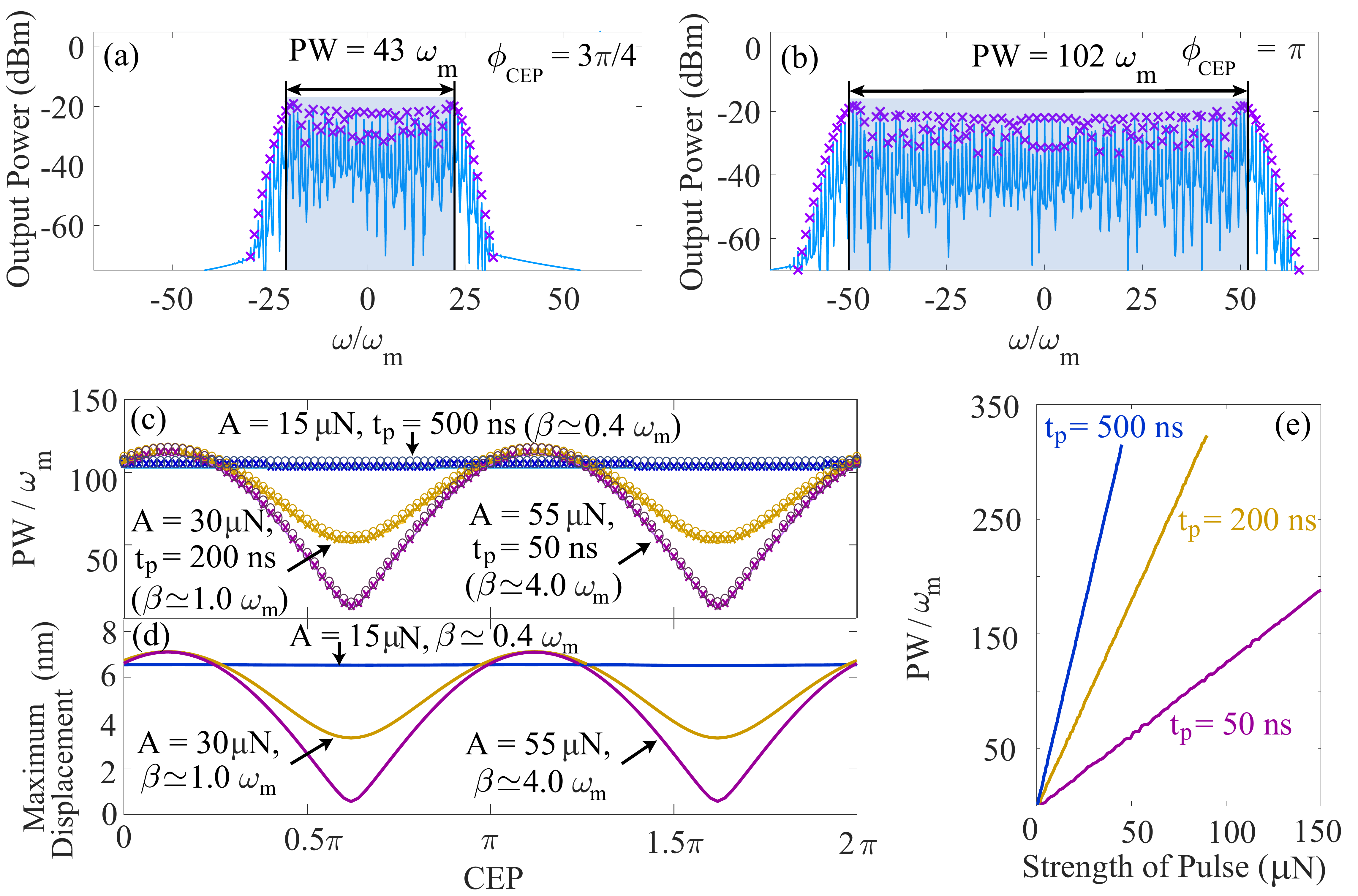} \caption{(a)--(b) The optical power spectrum obtained from the analytical solution of Eq.~(\ref{eq:opt_sp}) (cross marks), compared with the accurate numerical solution of Eqs.~(\ref{eq:dif1}) and (\ref{eq:dif2}) (lines). The relevant parameters are set to: $P_\textrm{l} = \SI{10}{\milli\watt}$, $A=\SI{200}{\micro\newton}$ and $t_\textrm{p} = \SI{20}{\nano\second}$. \;(c) The plateau width (PW) as a function of CEP. The lines and the cross marks indicate the PW obtained respectively from the numerical and analytical solutions, as before. The circle marks are obtained from the more concise definition of Eq.~(\ref{eq:pw}). \;(d) The CEP dependence of maximum mechanical displacement. \;(e) The numerical results for the linear relation of the PW to the amplitude of the input mechanical pulse.}
\label{fig:2}
\end{figure}

We start the analysis from the later time window, in which strong-drive and negligible-damping approximations are considered. The relatively small impact of the approximations is shown in the numerical results [see Fig.~\ref{fig:1}(c)] for the dynamics of the system driven by a pulse of even a micronewton. The spectrum for the cavity field is then given as follows (detailed calculations are available in the Appendix)
\begin{align}
c_{N}(\omega_\textrm{m})=\varepsilon_\textrm{l}e^{iN\phi_\textrm{m}}\sum_{n=-\infty}^{+\infty}\frac{J_{n}(\alpha)J_{N+n}(\alpha)}{i(n\omega_\textrm{m}-\Delta)+\kappa}, \label{eq:opt_sp}
\end{align}
where $\alpha=Gq_{0}/\omega_\textrm{m}$, $N$ symbolizes the order of the optical sidebands, and $J_{n}(\alpha)$ is the Bessel function of the first kind with $n$ for its integer order. The parameters $q_0$ and $\phi_\textrm{m}$ are the amplitude and the initial phase of the mechanical oscillation respectively, and their values are extracted from numerical results. Given the property of the Bessel function $\sum_{n=-\infty}^{+\infty}J_{n}(\alpha)J_{N+n}(\alpha)=\delta_{N0}$ (where $\delta_{N0}$ is the Kronecker delta function), it is shown from Eq.~(\ref{eq:opt_sp}) that $c_{N}(\omega_\textrm{m})(N\neq0)$ goes to zero when the cavity decay $\kappa$ is much larger than the mechanical frequency $\omega_\textrm{m}$. This means that the the higher-order sidebands and associated effects are absent in such a regime, and thus our following work will be focused only on the regime where $\kappa\ll\omega_\textrm{m}$, i.e., the resolved-sideband regime. From now on we choose $\Delta=-\omega_\textrm{m}$ and consider the power spectrum of the transmitted field to avoid the interference between the output field and the input field by applying the input-output relation: $c_{N}^{out}(\omega_m)=\sqrt{2\kappa}c_{N}(\omega_\textrm{m})$. The output power spectrum is shown in Fig.~\ref{fig:2}(a)-(b), in which a plateau is formed by the emergence of high-order sidebands~\cite{xiong_carrier-envelope_opto, cao_HOS-generation_opto} that ends swiftly at a specific cut-off frequency. 
The plateau width (PW) is a convenient quantity to consider for a characterization of the system's conditions. We define it from the difference between the highest and the lowest frequencies corresponding to the sidebands whose amplitudes are greater than half of the maximum value, determining the cut-off frequencies of the anti-Stokes and the Stokes sidebands respectively. When the mechanical pulse contains only a few cycles, the PW oscillates periodically with period $\pi$ as the value of CEP changes [see\ Fig.~\ref{fig:2}(c)]. This is a clear CEP-dependent effect that can be utilized to infer the shape of the pulse and prompt the necessary actions for its control. Figure~\ref{fig:2}(c) shows in particular that the CEP-dependent effect is weakened as the temporal width of the pulse, $t_\textrm{p}$, becomes larger (i.e. the pulse contains more cycles), and ultimately vanishes when $t_\textrm{p}$ is large enough. We note that the Bessel function, $J_{n}(\alpha)$, decays very quickly when $n>\alpha$. Applying this property to Eq.~(\ref{eq:opt_sp}), we see that higher order terms in the series reduce to zero rapidly once $N>\alpha$. This means that the cut-off frequencies in the output power spectrum can be approximated to $\pm\alpha\omega_\textrm{m}$. We can therefore redefine the PW $\varpi$ in a more concise equation:
\begin{align}
\varpi=2\alpha\omega_\textrm{m}=2Gq_{0}. \label{eq:pw}
\end{align}
As seen in the overlap of the PW traces in Fig. \ref{fig:2}(c), this new definition distinctly agrees with the previous one. Equation~(\ref{eq:pw}) reminds us that it is essential to fully understand the behaviors of the maximum displacement ($q_{0}$) for the sake of obtaining the physical picture of the CEP-dependent effect. For this reason, in the following discussions we focus on the analysis of the mechanical spectrum in the time window during which the pulse drives the system. 

By dropping the radiation pressure term as in the previous calculation and considering the Fourier transform of $q(t)$  [$q(t)=\int_{-\infty}^{+\infty}q(\omega)e^{i\omega t}d\omega$], we give the displacement spectrum (see Appendix):
\begin{align}
\left|q(\omega)\right|^2 = \left|\frac{A}{4\beta\sqrt{\pi}}\right|^2\frac{1}{\left|\chi(\omega)\right|^2}\left(e^{-\frac{(\omega-\omega_\textrm{m})^{2}}{2\beta^{2}}}+e^{-\frac{(\omega+\omega_\textrm{m})^{2}}{2\beta^{2}}}\right.\nonumber\\
+\qquad\qquad\left.2e^{-\frac{\omega^{2}+\omega_\textrm{m}^{2}}{2\beta^{2}}}\cos[2(\omega_\textrm{m}t_{0}+\phi_\textrm{CEP})]\right),
\label{eq:dis_a}
\end{align}
where $\chi(\omega)=[m(\omega_\textrm{m}^2-\omega^2+i\gamma_\textrm{m}\omega)]^{-1}$ is the susceptibility of the mechanical oscillator. This spectrum shows that the maximum values of the mechanical displacement are found at the mechanical sidebands, i.e. $\omega\approx\omega_\textrm{m}$ and $\omega\approx-\omega_\textrm{m}$. The term including the CEP in Eq.~(\ref{eq:dis_a}) reduces to $2\exp(-\omega_\textrm{m}^{2}/\beta^{2})\cos[2(\omega_\textrm{m}t_{0}+\phi_\textrm{CEP})]$ if one looks at these two mechanical sidebands. Since the PW is linearly related to the maximum mechanical displacement [see\ Eq.~(\ref{eq:pw})], such term indicates that both are modulated periodically by the CEP with period $\pi$ when the parameter $\beta$ is much greater than $\omega_\textrm{m}$ (i.e., $t_\textrm{p}\ll\sqrt{2\ln2}/\omega_\textrm{m}$). Moreover, the dependence of the PW (and maximum displacement) on the CEP fades with decreasing $\beta$ (increasing $t_\textrm{p}$), and such dependence completely dies out when $\beta\ll\omega_\textrm{m}$. These features agree very well with previous discussions about CEP-dependent effects (see\ Fig.~\ref{fig:2}). Furthermore, we find that the numerical result for the maximum mechanical displacement as a function of CEP [see\ Fig.~\ref{fig:2}(d)] shows the same dynamics as the analytical results. We therefore identify the regime where $\beta\gg\omega_\textrm{m}$ (pulse contains few cycle) as the \textit{few-cycle} regime and the regime where $\beta\ll\omega_\textrm{m}$ (pulse contains many cycle) as the \textit{multi-cycle} regime. We note from Eq.~(\ref{eq:dis_a}) that the mechanical displacement linearly depends on the amplitude of the mechanical pulse (i.e., $A$), implying that the PW is also proportional to $A$ [see\ Fig.~\ref{fig:2}(e)]. Such dependence provides a potential and concise way to linearly manipulate the line number of an optical comb. Since the line spacing of the optical comb observed in Fig.~\ref{fig:2}(a)--(b) is determined by the mechanical frequency (the value is taken from an experiment to be $2\pi\times947$ kHz~\cite{groblacher_strong-coupling-exp_opto}), it is in principle possible to beat the challenge of narrow-line spacing~\cite{jiang_shaping-comb_opulse}. Current experiments in optomechanics have achieved the mechanical frequencies lower than kilohertz \cite{aspelmeyer_cavity-opto_review, kleckner_experiment-trampoline_opto}, meaning that this technique would open access to new regimes of optical comb manipulation. 

Based upon the analysis above, we devise a two-step interpretation of the appearance of high-order sidebands induced in the OMS: (i) an intense mechanical pulse is sent to the mechanical resonator to drive high-amplitude oscillations; (ii) the strong oscillations of the mechanical resonator induce Stokes and anti-Stokes scattering of the cavity field. The requirement for the first step is embodied in Eq.~(\ref{eq:dis_a}), according to which the higher-order sidebands forming the plateau can be excited only when $\alpha\gg1$, or equivalently $q_{0}\gg\omega_\textrm{m}/G$. The second step is represented by Eq.~(\ref{eq:opt_sp}), describing how both the Stokes and the anti-Stokes fields build up the power spectrum. The two-step model can also be used to obtain a physical insight into the appearance of CEP-dependent effects. The first step introduces an intense mechanical pulse to drive the oscillations of the mechanical resonator. Since the amplitude of the oscillations depends on the CEP [see Eq.~(\ref{eq:dis_a}) and Fig.~\ref{fig:2}(b)], one can thus infer a similar dependence on the optical spectrum of the cavity, thanks to Eq.~(\ref{eq:pw}). The second step relates to the appearance of sidebands on the optical output spectrum due to the impact of the modulation from the mechanical oscillation. Observation of the PW of the output power spectrum will therefore give, thanks to Eq.~(\ref{eq:pw}), a direct measurement of the CEP.

\section{CEP-dependent effect in the multi-cycle regime}As mentioned earlier, traditional CEP-dependent effects are weakened in the presence of multiple cycles. In this section, we demonstrate how the inclusion of an auxiliary continuous optical field in the system lifts any requirement linked to the number of cycles for observing CEP-dependent effects. 
\begin{figure}[htb]
\includegraphics[width=0.49\textwidth]{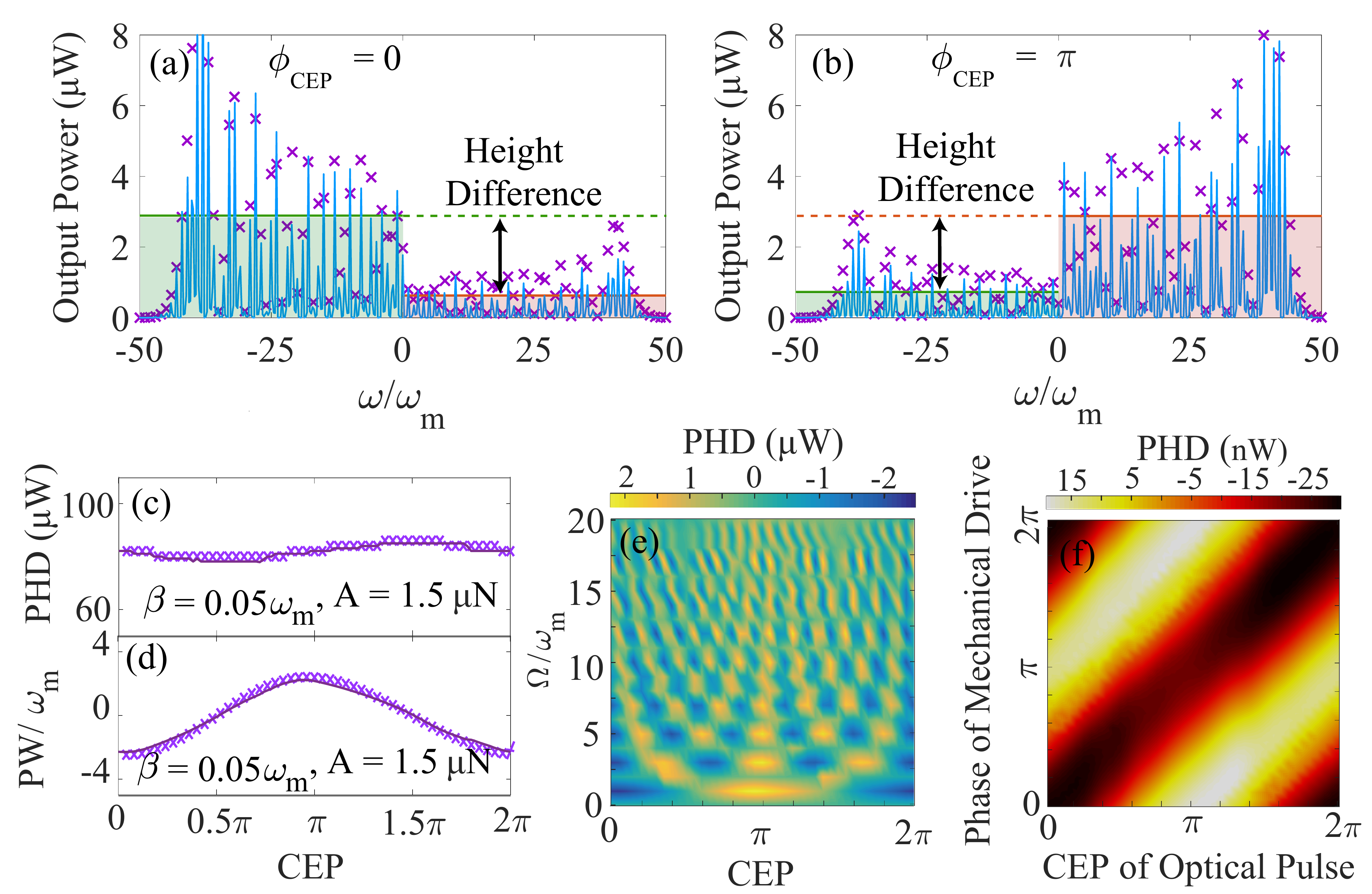} \caption{(a)--(b) The output power spectrum from Eq.~(\ref{eq:new_m}) (cross marks), compared with its  numerical results (lines) under no approximation. The parameters are set to: $P_\textrm{l} = \SI{2}{\milli\watt}$, $P_\textrm{a} = \SI{3}{\milli\watt}$, $A=\SI{6}{\micro\newton}$ and $\beta = 0.2\omega_m$. \;(c)--(d) Dependence of the plateau parameters (PW and PHD) on CEP. The input power of the two lasers are the same as panel (a) and (b). \;(e) PHD as a function of the CEP and the frequency of the auxiliary laser ($\Omega$). \;(f) For a pulsed \emph{optical} input in the multi-cycle regime, PHD is shown to depend on its CEP. The parameters are: $P_\textrm{l} = \SI{1}{\micro\watt}$, $P_\textrm{a} = \SI{100}{\nano\watt}$, $A=\SI{40}{\nano\newton}$ and $\beta = 0.005\omega_\textrm{m}$.}
\label{fig:3}
\end{figure}
We consider the following inputs to the OMS: a mechanical pulse containing a large number of cycles, one continuous red-detuned laser, and another continuous auxiliary laser of different wavelength. The Hamiltonian for the optical drives [see\ Eq.~(\ref{eq:Ha})] reads $H_\textrm{od}=i\hbar[(\varepsilon_\textrm{l}+\varepsilon_\textrm{a}e^{-i\Omega t-\phi_\textrm{o}})\hat{c}^{\dagger}-H.c.]$, where $\varepsilon_\textrm{a}$ is determined by the input power of the auxiliary laser using $\varepsilon_\textrm{a}=\sqrt{2\kappa P_\textrm{a}/(\hbar\omega_\textrm{a})}$, $\Omega$ is its optical frequency in the rotating frame of $\omega_\textrm{l}$, i.e. $\Omega = \omega_\textrm{a}-\omega_\textrm{l}$ (from here, we choose $\Omega=l\omega_\textrm{m}$, with $l=\pm 0,1,2,...$), and $\phi_\textrm{o}$ is the phase difference of the two lasers at $t=0$. Considering approximations and calculation methods similar to those in the previous section (see Appendix), we obtain the cavity field spectrum:
\begin{align}
c_{N}(\omega_{m})=e^{iN\phi_{m}}\sum_{n=-\infty}^{+\infty}\frac{[\varepsilon_\textrm{l}J_{n}(\alpha)+\varepsilon_{a}e^{-i(\phi_{a}-l\phi_{m})}J_{n+l}(\alpha)]J_{N+n}(\alpha)}{i(n+1)\omega_{m}+\kappa}. \label{eq:new_m}
\end{align}

The output power spectrum [see\ Fig.~\ref{fig:3}(a)-(b) with $l=1$] for such setup displays a striking asymmetric feature: the sidebands from Stokes scattering differ in intensity from the sidebands due to anti-Stokes processes. Such effect cannot be found in the absence of the auxiliary laser [see\ Fig.~\ref{fig:2}], and has the very important quality of depending on the CEP. Therefore, in this multi-cycle regime where the dependence of PW on CEP vanishes [see\ Fig.~\ref{fig:3}(c)], we can still define a parameter that identifies and characterizes the CEP: the \textit{plateau height difference} (PHD) between the average height of the anti-Stokes and the Stokes sidebands in the output power spectrum. In contrast to PW, which is periodic by $\pi$ [see\ Fig.~\ref{fig:2}(e)], the PHD reveals a dependence periodic by $2\pi$ [see Fig.~\ref{fig:3}(d) and Eq.~(\ref{eq:new_m})]. Equation~(\ref{eq:new_m}) offers a clear physical picture for such an effect. The origin of the PHD lies in the interference between the two optical fields. As the auxiliary laser enters the cavity, it acquires high-order sidebands due to the interaction with the mechanical motion, similarly to the original field with amplitude $\varepsilon_\textrm{l}$. The Stokes and anti-Stokes sidebands from the two lasers interfere unevenly, leading to the difference in amplitude between the two. Furthermore, Eq.~(\ref{eq:new_m}) also reveals that the CEP dependence period would be $2\pi/l$ if the frequency of the auxiliary laser is $l\omega_\textrm{m}$, as demonstrated by the numerical results shown in Fig.~\ref{fig:3}(e).

\section{From mechanical to optical pulses}To complete our analysis we explore the effect of multi-cycle \emph{optical} pulses, as opposed to mechanical pulses, as considered so far. We take, as inputs, a continuous high-amplitude mechanical drive, a continuous red-detuned laser, and an optical pulse. The Hamiltonians describing the drives are 
$H_\textrm{od}=i\hbar[(\varepsilon_\textrm{l}+\varepsilon_\textrm{a}e^{-i\Omega t-\phi_\textrm{CEP}}e^{-\beta^{2}(t-t_{0})^{2}})\hat{c}^{\dagger}-H.c.]$ and $H_\textrm{md}=A\cos(\omega_\textrm{m}t+\phi)\hat{q}$.

Like in the scenario for the mechanical pulse, the output power spectrum shows again the emergence of a dependence of the power spectrum on the CEP in the form of a PHD [see\ Fig.~\ref{fig:3}(f)], even in the regime where the optical pulse contains many cycles and traditional CEP-dependent effects are washed out. It should be noted that, additionally, traditional CEP-dependent effects in an ion platform can be observed only when the optical pulse is intense enough so that the tunneling ionization can take place~\cite{nakajima_CEP-weak_opulse}. In the OMS, this condition is generally transferred onto the drive for the mechanical resonator, which is required to be intense to generate a strong mechanical oscillation leading to higher-order sidebands. In a previous relevant work, the optical pulse was required to be intense to meet such a requirement~\cite{xiong_carrier-envelope_opto}. Here, it is the continuous mechanical drive that takes the responsibility to intensely drive the mechanical resonator. The new CEP measurement for the optical pulse based upon an interferometric process is therefore technically free from any intensity requirements.

\section{Conclusion}We extended the observation of CEP-dependent effects from ultra-fast optics to the realm of mechanical pulses using optomechanics. Approximated analytical solutions robustly supported by accurate numerical solutions were delivered, revealing a clear physical picture of the model. We developed a two-step model to describe the physical processes linking the optical spectrum to the CEP of few-cycle mechanical pulses, and then identified a novel effect that delivers information on the CEP regardless of how many cycles the envelope of the mechanical pulse contains. Importantly, such a method also functions even for ultra-weak optical pulses. The method described applies extensively to a variety of pulses: mechanical or optical, in the few- or multi-cycle regime, and without specific requisites for optical intensity. The diverse development of an experimental optomechanical setup \cite{aspelmeyer_cavity-opto_review} may enable our scheme to be implemented for the measurement of the CEP and manipulation of optical combs in a very wide frequency range, and the versatility of optomechanical platforms could be used to stretch these advantages to unexplored grounds (such as ultrasonic pulses~\cite{wright_pico-wide-frequency_sound}).

\begin{acknowledgments}
This research was funded by the Australian Research Council Centre of Excellence CE110001027 and the Australian Government Research Training Program Scholarship. PKL acknowledges support from the ARC Laureate Fellowship FL150100019. 
\end{acknowledgments}

\bibliographystyle{apsrev4-1}
\bibliography{mybib}

\appendix
\section{Mechanical pulses in the few-cycle regime}
We consider a cavity OMS driven by a continuous optical input with frequency $\omega_\textrm{l}$ and an external mechanical pulse with carrier frequency $\omega_\textrm{m}$, and its Hamiltonian in the rotating frame of $\omega_\textrm{l}$ is written as
\begin{gather}
H=(\frac{\hat{p}^{2}}{2m}+\frac{m\omega_{m}^{2}\hat{q}^{2}}{2})+\hbar\Delta\hat{c}^{\dagger}\hat{c}+\hbar G\hat{q}\hat{c}^{\dagger}c+H_{md}+H_{od}, \nonumber \\
\label{equ:Ha}
\end{gather}
with
\begin{gather}
H_{md}=A\exp[-2\ln2(\frac{t-t_{0}}{t_{p}})^{2}]\cos(\omega_{m}t+\phi)\hat{q},\\
H_{od}=i\hbar(\varepsilon_{l}\hat{c}^{\dagger}-H.c.).\label{equ:Ha-1}
\end{gather}

In this work, the classical Langevin equations are given as
follows:

\begin{gather}
\ddot{q}+\gamma_{m}\dot{q}+\omega_{m}^{2}q=-\frac{\hbar g}{m}\left\vert c\right\vert ^{2}+\frac{A}{m}e^{-\beta^2\left(t-t_{0}\right)^{2}}\cos(\omega_{m}t+\varphi), \label{dif1}\\
\dot{c}=-[\kappa+i(-\Delta+gq)]c+\varepsilon_{l}. \label{dif2}
\end{gather}

The approximated analytical solution of the two nonlinear differential equations can be calculated if we consider a \textit{strong-drive approximation}, where the drive to the mechanical motion is typically much more intense than the effect of the radiation pressure force, and thus we are able to drop the term $-\frac{\hbar G}{m}\left|c\right| ^{2}$ from Eq.~(\ref{dif1}). If we focus on the time window that starts right after the end of the mechanical pulse and ends before the damping of the displacement becomes appreciable, Eq.(\ref{dif1}) become $\ddot{q}+\omega_{m}^{2}q=0$, and its solution is easily obtained as follows: 
\begin{eqnarray}
q(t) & = & q_{0}\cos(\omega_{m}t+\phi_{m}),
\end{eqnarray}
where $q_0$ and $\phi_\textrm{m}$ are the amplitude and the initial phase of the mechanical oscillation respectively. We define $h(t)=-[\kappa+i(-\Delta+Gq)]$ and suppose the solution of $c(t)$ as follows,

\begin{eqnarray}
c(t) & = & e^{\int h(t)dt}g(t)\nonumber\\
 & = & f(t)g(t),
\end{eqnarray}
with $f(t) = e^{\int h(t)dt}$. Substituting this equation into Eq. (\ref{dif2}), we have
\begin{eqnarray}
f(t)g(t)h(t)+f(t)\dot{g}(t) & = & h(t)f(t)g(t)+\varepsilon_{l}, \nonumber\\
g(t) & = & \varepsilon_{l}\int f(t)^{-1}dt.
\end{eqnarray}

We give $f(t)$ as follows

\begin{eqnarray}
f(t) & = & e^{\int h(t)dt} \nonumber\\
 & = & \exp(-\kappa t)\exp(-i\omega_{m}t)\exp[-i\alpha\sin(\omega_{m}t+\phi_{m})], \nonumber
\end{eqnarray}

where $\alpha=\frac{Gq_{0}}{\omega_{m}}$. Then $g(t)$ is calculated as:

\begin{eqnarray}
g(t) & = & \varepsilon_{l}\int f(t)^{-1}dt \nonumber\\
 & = & \varepsilon_{l}\int\exp(\kappa t)\exp(i\omega_{m}t)\{\cos[\alpha\sin(\omega_{m}t+\phi_{m})]\nonumber\\
 &  & +i\sin[\alpha\sin(\omega_{m}t+\phi_{m})]\}dt\nonumber\\
 & = & \varepsilon_{l}\int\exp(\kappa t+i\omega_{m}t)\nonumber\\
 & &\{\frac{1}{2}\sum_{n=-\infty}^{+\infty}[J_{n}(\alpha)e^{in(\omega_{m}t+\phi_{m})}+J_{n}(\alpha)e^{-in(\omega_{m}t+\phi_{m})}]\nonumber\\
 &  & -\frac{1}{2}\sum_{n=-\infty}^{+\infty}[-J_{n}(\alpha)e^{in(\omega_{m}t+\phi_{m})}+J_{n}(\alpha)e^{-in(\omega_{m}t+\phi_{m})}]\}dt\nonumber\\
 & = & \varepsilon_{l}\sum_{n=-\infty}^{+\infty}J_{n}(\alpha)e^{in\phi_{m}}\frac{\exp[i(n+1)\omega_{m}t+\kappa t]}{i(n+1)\omega_{m}+\kappa}.
\end{eqnarray}

We thus obtain $c(t)$:

\begin{eqnarray}
c(t) & = & f(t)g(t)\nonumber\\
 & = & \varepsilon_{l}\sum_{k=-\infty}^{+\infty}\sum_{n=-\infty}^{+\infty}J_{n}(\alpha)J_{k}(\alpha)e^{i(n-k)\phi_{m}}\frac{\exp[i(n-k)\omega_{m}t]}{i(n+1)\omega_{m}+\kappa}\nonumber\\
 & = & \sum_{N=-\infty}^{+\infty}c_{N}(\omega_{m})\exp[-iN\omega_{m}t],
\end{eqnarray}
with
\begin{eqnarray}
c_{N}(\omega_{m})=\varepsilon_{l}e^{iN\phi_{m}}\sum_{n=-\infty}^{+\infty}\frac{J_{n}(\alpha)J_{N+n}(\alpha)}{i(n+1)\omega_{m}+\kappa}.
\end{eqnarray}

Subsequently, we calculate the spectrum of the mechanical displacement for the time window during which the pulse is interacting with the system. By dropping the radiation pressure term like in the previous calculation, Eq.~(\ref{dif1}) takes the form:

\begin{equation}
\ddot{q}+\gamma_{m}\dot{q}+\omega_{m}^{2}q=\frac{A}{m}e^{-\beta^2\left(t-t_{0}\right)^{2}}\cos(\omega_{m}t+\varphi),\label{mechanical}
\end{equation}

The right side of the equation above is expanded as follows:

\begin{widetext}
\begin{eqnarray}
\ddot{q}+\gamma_{m}\dot{q}+\omega_{m}^{2}q & = & \frac{A}{m}e^{-\beta^{2}(t-t_{0})^{2}}\cos\text{(}\omega_{m}t+\varphi)\nonumber\\
 & = & \frac{A}{m}\int_{-\infty}^{+\infty}\frac{1}{2\beta\sqrt{\pi}}\exp(-\frac{\omega^{2}}{4\beta^{2}})\exp[i\omega(t-t_{0})][\frac{e^{i\omega_{m}t+i\varphi}+e^{-i\omega_{m}t-i\varphi}}{2}]d\omega\nonumber\\
 & = & \frac{A}{m}\int_{-\infty}^{+\infty}\frac{1}{4\beta\sqrt{\pi}}\exp[-\frac{(\omega-\omega_{m})^{2}}{4\beta^{2}}]e^{-i(\omega t_{0}-\omega_{m}t_{0}-\varphi)}e^{i\omega t}d\omega\nonumber\\
 &  & +\frac{A}{m}\int_{-\infty}^{+\infty}\frac{1}{4\beta\sqrt{\pi}}\exp[-\frac{(\omega+\omega_{m})^{2}}{4\beta^{2}}]e^{-i(\omega t_{0}+\omega_{m}t_{0}+\varphi)}e^{i\omega t}d\omega. \label{equ:q_pulse}
\end{eqnarray}
\end{widetext}

We substitute the ansatz $q(t)=\int_{-\infty}^{+\infty}q(\omega)e^{i\omega t}d\omega$
into the left side of the Eq. (\ref{mechanical}), and obtain:

\begin{equation}
m\ddot{q}+m\gamma_{m}\dot{q}+m\omega_{m}^{2}q=\int_{-\infty}^{+\infty}q(\omega)(-\omega^{2}+i\gamma_{m}\omega+\omega_{m}^{2})e^{i\omega t}d\omega.\label{equ:q_ansatz}
\end{equation}

Comparing Eqs. (\ref{equ:q_pulse}) with (\ref{equ:q_ansatz}), $q(\omega)$ is easily given as follows:

\begin{equation}
q(\omega)=\frac{A}{4m\beta\sqrt{\pi}}e^{-i\omega t_{0}}\frac{e^{-\frac{(\omega-\omega_{m})^{2}}{4\beta^{2}}}e^{i(\omega_{m}t_{0}+\varphi)}+e^{-\frac{(\omega+\omega_{m})^{2}}{4\beta^{2}}}e^{-i(\omega_{m}t_{0}+\varphi)}}{-\omega^{2}+i\gamma_{m}\omega+\omega_{m}^{2}}, \nonumber
\end{equation}

\begin{align}
\left|q(\omega)\right|^2 & = \left|\frac{A}{4\beta\sqrt{\pi}}\right|^2\frac{1}{\left|\chi(\omega)\right|^2}\left(e^{-\frac{(\omega-\omega_\textrm{m})^{2}}{2\beta^{2}}}+e^{-\frac{(\omega+\omega_\textrm{m})^{2}}{2\beta^{2}}}\right.\nonumber\\
&+\qquad\qquad\left.2e^{-\frac{\omega^{2}+\omega_\textrm{m}^{2}}{2\beta^{2}}}\cos[2(\omega_\textrm{m}t_{0}+\phi_\textrm{CEP})]\right),
\label{eq:dis_a}
\end{align}
where $\chi(\omega)=[m(\omega_\textrm{m}^2-\omega^2+i\gamma_\textrm{m}\omega)]^{-1}$ is the susceptibility of the mechanical oscillator.

\section{Mechanical pulses in the multi-cycle regime}
Here, we introduce another laser to release the few-cycle restriction. Similarly, the Hamiltonian of the drives can be written as
\begin{gather}
H_{md}=A\exp[-2\ln2(\frac{t-t_{0}}{t_{p}})^{2}]\cos(\omega_{m}t+\phi)\hat{q},\\
H_{od}=i\hbar[(\varepsilon_{l}+\varepsilon_{a}e^{-i\Omega t-i\phi_{a}})\hat{c}^{\dagger}-H.c.].\label{equ:Ha-1-2-1}
\end{gather}

We give the equations of motion as follows:

\begin{gather}
\ddot{q}+\gamma_{m}\dot{q}+\omega_{m}^{2}q=-\frac{\hbar g}{m}\left\vert c\right\vert ^{2}+\frac{A}{m}e^{-\beta^2\left(t-t_{0}\right)^{2}}\cos(\omega_{m}t+\phi), \label{mdif1}\\
\dot{c}=-[\kappa+i(-\Delta+gq)]c+(\varepsilon_{l}+s_{a}e^{-i\Omega t-i\phi_{a}}). \label{mdif2}
\end{gather}

When the interaction between the pulse and the system ends, Eq. (\ref{mdif1}), under the strong-drive and negligible-damping approximations, can be reduced to the following form as previous section:

\begin{eqnarray}
q(t) & = & q_{0}\cos(\omega_{m}t+\phi_{m}).
\end{eqnarray}

Like in Appendix A, we define $h(t)=-[\kappa+i(-\Delta+Gq)]$ and suppose the solution of $c(t)$ as follows,

\begin{eqnarray}
c(t) & = & e^{\int h(t)dt}g(t) \nonumber\\
 & = & f(t)g(t).
\end{eqnarray}

Substituting this equation into Eq. (\ref{mdif2}), we can obtain
\begin{eqnarray}
f(t)g(t)h(t)+f(t)\dot{g}(t) & = & h(t)f(t)g(t)+\sqrt{\eta_{c}\kappa}(\varepsilon_{l}+s_{a}e^{-i\Omega t-i\phi_{a}}),\nonumber\\
g(t) & = & g_{1}(t)+g_{2}(t),\nonumber
\end{eqnarray}

where 
\begin{eqnarray}
g_{1}(t)=\varepsilon_{l}\int f(t)^{-1}dt, \nonumber\\
g_{2}(t)=s_{a}e^{-i\phi_{a}}\int e^{-i\Omega t}f(t)^{-1}dt. \nonumber
\end{eqnarray}

We give $f(t)$ and $g_{1}(t)$ as follows, 

\begin{eqnarray}
f(t) & = & e^{\int h(t)dt}\nonumber\\
 & = & \exp(-\kappa t)\exp(-i\omega_{m}t)\exp[-\frac{iGq_{0}}{\omega_{m}}\sin(\omega_{m}t+\phi_{m})]\nonumber\\
 & = & \exp(-\kappa t)\exp(-i\omega_{m}t)\sum_{n=-\infty}^{+\infty}J_{n}(\alpha)e^{-in(\omega_{m}t+\phi_{m})}dt, \nonumber\\
g_{1}(t) & = & \varepsilon_{l}\int f(t)^{-1}dt\nonumber\\
 & = & \varepsilon_{l}\sum_{n=-\infty}^{+\infty}J_{n}(\alpha)e^{in\phi_{m}}\frac{\exp[i(n+1)\omega_{m}t+\kappa t]}{i(n+1)\omega_{m}+\kappa}, \nonumber
\end{eqnarray}
where $\alpha=\frac{Gq_{0}}{\omega_{m}}$. 

Similarly, $g_{2}(t)$ is obtained as
\begin{eqnarray}
g_{2}(t) & = & \varepsilon_{a}e^{-i\phi_{a}}\int e^{-il\omega_{m}t}f(t)^{-1}dt  \nonumber\\
& = &\varepsilon_{a}e^{-i\phi_{a}}\sum_{n=-\infty}^{+\infty}J_{n+l}(\alpha)e^{i(n+l)\phi_{m}}\frac{\exp[i(n+1)\omega_{m}t+\kappa t]}{i(n+1)\omega_{m}+\kappa}. \nonumber
\end{eqnarray}
As a result, we obtain $c(t)$:
\begin{eqnarray}
c(t) & = & f(t)(g_{1}(t)+g_{2}(t)),\nonumber
\end{eqnarray}
with
\begin{eqnarray}
c_{N}(\omega_{m})&=&e^{iN\phi_{m}}\sum_{n=-\infty}^{+\infty}\frac{[\varepsilon_{l}J_{n}(\alpha)+\varepsilon_{a}e^{-i(\phi_{a}-l\phi_{m})}J_{n+l}(\alpha)]J_{N+n}(\alpha)}{i(n+1)\omega_{m}+\kappa}. \nonumber
\end{eqnarray}

\end{document}